\documentclass{article}

\usepackage{url}
\usepackage{xcolor}
\usepackage{amsmath}
\usepackage{array}
\usepackage{multirow}

\usepackage{graphicx}  

\usepackage{enumitem}  
\usepackage{hyperref}
\usepackage{arxiv}
\usepackage{amsmath}    
\usepackage{amssymb}    
\usepackage{enumitem}   
\usepackage{graphicx}   
\usepackage{longtable}  
\usepackage{parskip}    
\usepackage{array}      
\usepackage{booktabs}   
\usepackage{hyperref}   
\usepackage{geometry}   
\usepackage{fancyhdr}   
\usepackage{tikz}       
\usepackage{caption}    

\usepackage[utf8]{inputenc} 
\usepackage[T1]{fontenc}    
\usepackage{hyperref}       
\usepackage{url}            
\usepackage{booktabs}       
\usepackage{amsfonts}       
\usepackage{nicefrac}       
\usepackage{microtype}      
\usepackage{lipsum}
\usepackage{graphicx}
\graphicspath{ {./images/} }

\title{Dual-Alignment Knowledge Retention for Continual Medical Image Segmentation}

\author{
 Yuxin Ye \\
Sichuan University\\
   \And
 Yan Lu \\
  Sichuan University\\
  \And
 Shujian Yu \\
  Vrije Universiteit Amsterdam\\
}

\begin{document}
\maketitle
\begin{abstract}
Continual learning in medical image segmentation involves sequential data acquisition across diverse domains (e.g., clinical sites), where task interference between past and current domains often leads to catastrophic forgetting. Existing continual learning methods fail to capture the complex dependencies between tasks. We introduce a novel framework that mitigates forgetting by establishing and enhancing complex dependencies between historical data and the network in the present task. Our framework features a dual-alignment strategy, the cross-network alignment (CNA) module aligns the features extracted from the bottleneck layers of the current and previous networks, respectively, while the cross-representation alignment (CRA) module aligns the features learned by the current network from historical buffered data and current input data, respectively. Implementing both types of alignment is a non-trivial task. To address this, we further analyze the linear and nonlinear forms of the well-established Hilbert-Schmidt Independence Criterion (HSIC) and deliberately design feature mapping and feature pairing blocks within the CRA module. Experiments on medical image segmentation task demonstrate our framework’s effectiveness in mitigating catastrophic forgetting under domain shifts. 
\end{abstract}

\section{Introduction}
\label{sec:intro}
Medical image segmentation is a key technology used to automatically or semi-automatically divide anatomical structures or lesions in images. This helps doctors accurately locate disease areas and assess treatment outcomes  \cite{r1.1,r1.2}. However, medical image data are often temporal in nature, originating from different medical institutions or imaging devices, and tend to accumulate over time. The diversity of data sources makes continuous learning challenging. Traditional sequential training \cite{r1.3} can lead to catastrophic forgetting, where models lose previously learned information when exposed to new data.

Continual learning \cite{r1.5,r1.6,r1.7} enables models to retain knowledge from previous tasks while acquiring new information, making it particularly valuable for medical image segmentation, where data are diverse and evolve over time. Domain continual learning \cite{r1.8,r1.9,r1.10}, a specialized form of continual learning, enhances model adaptability to cross-domain variations, improving cross-platform robustness. Additionally, it facilitates real-time adaptation to evolving patient data, aiding doctors in refining treatment plans. In personalized medicine, continual learning offers a flexible approach for managing individual patient data, ensuring more adaptive and tailored healthcare solutions.

There are different approaches to address the issue of
forgetting in continual learning. Replay-based methods \cite{r2.7,r2.8,r2.9} store subsets of past task data in a memory buffer and interleave them with new task samples during training. Regularization approaches \cite{r2.4,r2.5} mitigate interference by penalizing changes to parameters critical for prior tasks, thereby preserving historical knowledge. In contrast, parameter isolation techniques \cite{r2.2,r2.3} assign dedicated network parameters to individual tasks, eliminating overlap and minimizing cross-task interference. Lastly, knowledge distillation \cite{rcross_organ1,r2.22} transfers learned information from a trained teacher model to a student model, enabling the student to incrementally adapt to new tasks while retaining essential features from earlier ones.

Existing continual learning approaches struggle to fully capture the complex dependencies between consecutive tasks, making them inadequate for addressing catastrophic forgetting. To overcome this limitation, we propose an innovative framework that constructs and enhances structured relationships between the current task and previously acquired knowledge. Our framework incorporates a dual-alignment strategy with Cross-Network Alignment (CNA) and Cross-Representation Alignment (CRA). The CNA module aligns features extracted from the bottleneck layers of the current and previously learned networks, ensuring that the current network's behavior closely resembles that of its predecessor and thereby facilitating knowledge retention at the network level. The CRA module maximizes the dependence between bottleneck-layer features learned by the current network from historical buffered data and current input data, thereby promoting knowledge retention at the feature level. 

Unfortunately, implementing CRA presents two major challenges. First, the high-dimensional nature of 3D tensor latent representations complicates the alignment process. Second, the absence of explicit sample correspondence further hinders effective feature pairing. To address these challenges, we propose a Feature Mapping (FM) mechanism that compresses 3D feature maps into compact vector representations, reducing dimensionality to enable more efficient alignment. In addition, a Feature Pairing (FP) strategy leverages nonlinear Hilbert-Schmidt Independence Criterion (HSIC) \cite{Gretton} to maximize mutual information across tasks through exhaustive pairing, thereby enhancing knowledge transfer.
By integrating CNA and CRA, our framework establishes bidirectional dependencies between current tasks and historical models, facilitating efficient knowledge transfer while preserving domain-invariant features. Our main contributions in this work include:
\begin{itemize}[label=\textbullet]
\item We introduce an innovative dual-alignment framework that establishes dependencies between historical and current data network features, effectively leveraging their information to reduce catastrophic forgetting.
    
\item To align representations in CRA module, we design a feature mapping block followed by a feature pairing block. Within the CNA module, we examine the linear form of HSIC and introduce a computationally efficient surrogate to facilitate alignment.

\item Experiments on representative medical image datasets against $8$ state-of-the-art (SOTA) methods validate the effectiveness and superiority of ours.

\end{itemize}


\begin{figure*}[t]  
    \centering
    \includegraphics[width=0.98\textwidth, trim=0 0 0 0, clip]{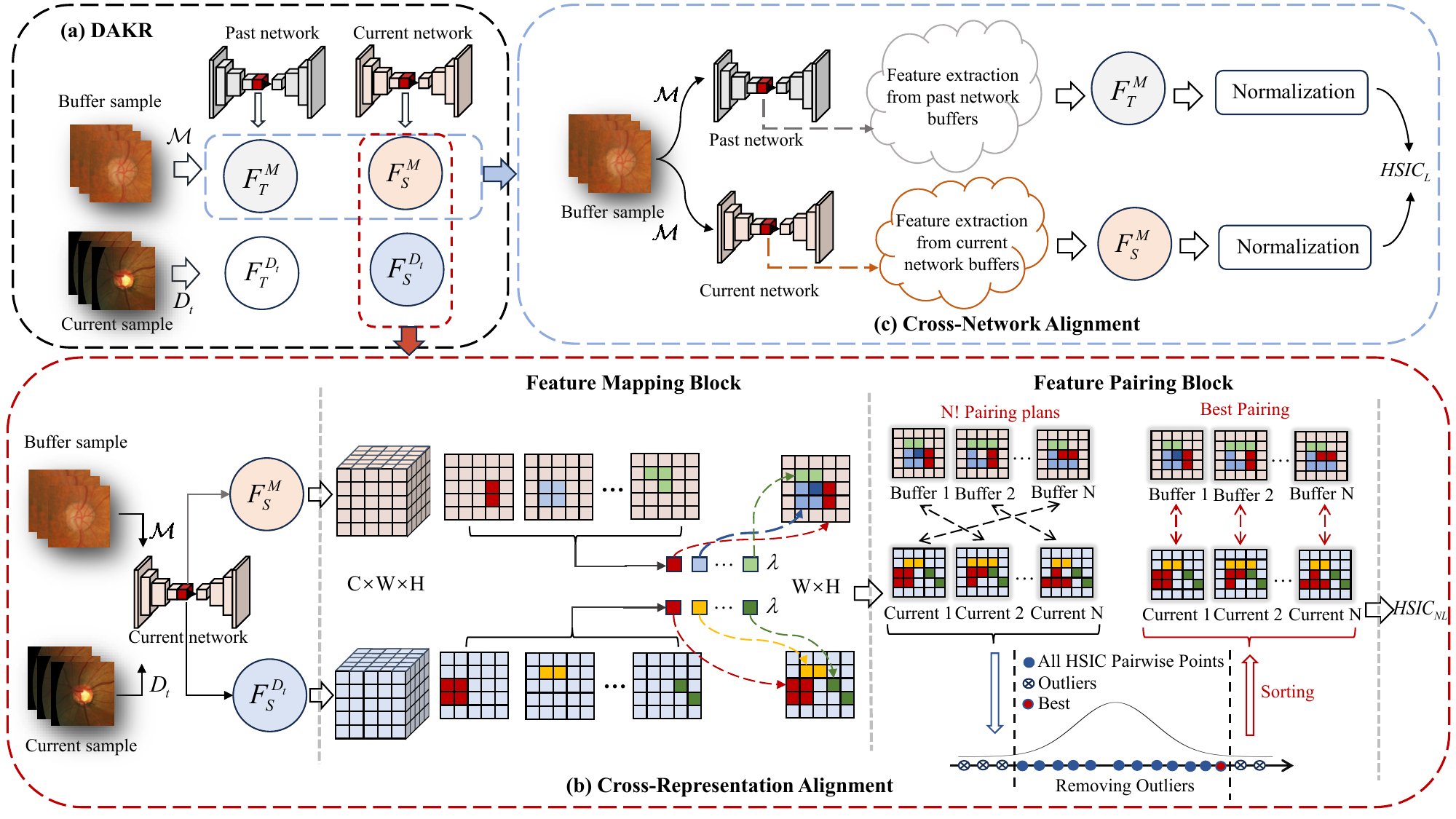}
    \vspace{-0.2cm}
    \caption{The framework of our DAKR-HSIC: (a) the overall architecture, (b) the Cross-Representation Alignment (including Feature Mapping and Feature Pairing); and (c) the Cross-Network Alignment, all working together to enable effective knowledge transfer for domain lifelong medical image segmentation. The cursive $\mathcal{M}$ denotes the buffer of past samples, $D_{t_c}$ the current sample, and $F$ refers to latent feature representations.}
    \label{figure1}
\end{figure*}

\section{Related Work and Preliminaries}
\subsection{Continual Medical Image Segmentation}

Continual learning in medical imaging involves updating the underlying model with new data while retaining previous knowledge. Current research primarily focuses on two learning scenarios: cross-organ~\cite{rcross_organ1,rcross_organ2,rcross_organ3} and cross-domain. Cross-organ learning targets segmentation across different organs, while cross-domain learning typically focuses on images of the same organ acquired using different medical equipment, aiming to address variations introduced by different clinical sites or imaging modalities.
 
Cross-domain learning is highly valuable in practice because deep learning models are often developed and deployed incrementally over time. For instance, a model trained and validated on data from a single healthcare institution (e.g., Hospital-A) is subsequently disseminated and implemented at other sites (e.g., Hospital-B, Hospital-C, etc.), often as the business or healthcare network expands~\cite{amrollahi2022leveraging}.
Although there has been some emerging research on domain continual learning in medical imaging, the focus has predominantly been on classification tasks \cite{r2.23,r2.24,r2.25,r2.26}. Our paper deals with the domain continual medical image segmentation. 

Memory replay helps retain past knowledge by storing samples from previous tasks. \cite{r2.13} propose a replay module that efficiently recreates past data while minimizing reliance on domain-specific features. \cite{r2.14} introduce a memory bank that is constructed by selecting images that make significant contributions to learning. \cite{r2.15} utilize dynamic memory, enabling the model to retain old data and balance new and old information by identifying style clusters within the data stream.

Regularization strategies retain old task knowledge by constraining parameter updates. MAS-LR \cite{r2.19} and \cite{r2.20} adjust learning rates and use Fisher information to consolidate key weights. \cite{r2.16} introduce a selective regularization approach that protects key knowledge through shape and semantic awareness. Lifelong nnU-Net \cite{r2.17} use regularization to ensure stability. \cite{r2.18} address this by introducing a low-rank expert mixture, reducing task interference. 

Recently, knowledge distillation has gained attention in continual learning for medical image segmentation, where it transfers knowledge from teacher to student models to mitigate catastrophic forgetting. For example, 
\cite{r2.22} propose a tri-enhanced distillation framework that improves knowledge redundancy reduction, selective transfer, and bias reduction during knowledge fusion. However, relying solely on uncertainty for fusion may overemphasize certain domain knowledge, limiting its ability to address feature differences between domains. Noise and blurring artifacts in medical images also complicate uncertainty estimation, affecting decision accuracy in critical areas.

\subsection{Hilbert-Schmidt Independence Criterion}

HSIC is a method used to measure the statistical dependence between two random variables. While alternative methods like Mutual Information (MI) offer insight into dependence, they are challenging to estimate in high-dimensional spaces. The widely used $k$NN estimator for MI is not differentiable, and the Mutual Information Neural Estimator (MINE)~\cite{belghazi2018mutual} requires an auxiliary network, often resulting in training instability. In contrast, HSIC provides an elegant closed-form expression, is scalable to high-dimensional data, and avoids the need for additional networks, offering superior performance without the training complexities of methods like MINE.

For the nonlinear case, HSIC relies on kernel functions $k\left(x, x^{\prime}\right)$ and $l\left(y, y^{\prime}\right)$, with the empirical estimator given by:
\begin{equation}\label{eq:hsic_nonlinear}
\widehat{\mathrm{HSIC}}_{\text {nonlinear }}\left(\left\{x_i, y_i\right\}_{i=1}^N\right)=\frac{1}{N^2} \operatorname{tr}(K J L J) ,
\end{equation}
where, $K$ and $L$ are kernel matrices constructed using nonlinear kernel functions, and $J$ is the centering matrix, which allows capturing the complex nonlinear relationships within the data.

For the linear case, the feature mappings are $\phi(x)=x$ and $\psi(y)=y$, and the empirical estimator is:
\begin{equation}
\widehat{\operatorname{HSIC}}_{\text {linear }}\left(\mathcal{F}, \mathcal{G},\left\{x_i, y_i\right\}_{i=1}^N\right)=\frac{1}{N^2} \operatorname{tr}\left(X X^{\top} Y Y^{\top}\right) ,
\end{equation}
This simplifies to:
\begin{equation}
\label{eq:hsic_linear}
\frac{1}{N^2} \operatorname{tr}\left(X^{\top} Y\left(X^{\top} Y\right)^{\top}\right)=\frac{1}{N^2}\left\|X^{\top} Y\right\|_F^2=\|C\|_F^2 ,
\end{equation}
where $C$ is the empirical cross-correlation matrix and $\|\cdot\|_F$ is the Frobenius norm. Note that, since $X$ and $Y$ are standardized, the cross-covariance matrix becomes the cross-correlation matrix, where $(C)_{ij}\in [-1,1], \forall 1\leq i\leq d_X, 1\leq j\leq d_Y $.

Linear HSIC simplifies computation by leveraging the diagonal elements of the sample covariance matrix, making it suitable for capturing simple dependencies. Nonlinear HSIC, which involves complex kernel functions, is capable of capturing more intricate dependency structures.

\section{Method}
\label{sec:Method}
\subsection{Overview}
We propose HSIC-based Dual-Alignment Knowledge Retention (DAKR-HSIC) for domain continual medical image segmentation by establishing dynamic dependencies among current tasks, historical data, and network properties. As shown in Fig.~\ref{figure1}(a) Our solution features a dual-alignment strateg: a Cross-Representation Alignment (CRA) module and Cross-Network Alignment (CNA). The CRA module, depicted in Fig.~\ref{figure1}(b), is designed to extract task-invariant knowledge, with the motivation similar to the invariant representation learning in domain generalization. By maximizing the dependence between representations learned by the student network from both buffer data and the current task, we promote effective knowledge transfer and reduce forgetting. The CNA module, illustrated in Fig.~\ref{figure1}(c), encourages the new network to retain knowledge learned by the old network within the latent representation space. 
 
The overall objective of DAKR-HSIC is defined as follows:
\begin{equation}
\begin{split}
L_{\text{DAKR}} = & L_{\text{seg}} + \lambda_{\text{1}} L_{\text{REKD}} + \lambda_{\text{2}} L_{\text{CRA}} + \lambda_{\text{3}} L_{\text{CNA}}, 
\end{split}
\label{eq1}
\end{equation}
where $L_{\text {seg }}$ is the segmentation loss in current task, $L_{\text{REKD}}$ is a standard Knowledge distillation loss defined over both buffered data and the current task, with details presented in Section~\ref{sec:3.2}. $L_{\text{CRA}}$ and $L_{\text{CNA}}$ represent the CRA and CNA loss, respectively, both of which can be efficiently estimated or approximated by HSIC, as discussed in Section~\ref{sec:3.3}. The terms $\lambda_{1}, \lambda_{\text {2}}$ and $\lambda_{\text {3}}$ are positive regularization coefficients.

\subsection{Replay-Enhanced Knowledge Distillation} \label{sec:3.2}

Suppose we have $T$ domains, with each domain $t$ having its specific images $x$ and corresponding segmentation labels $y$, drawn \emph{i.i.d.} from an unknown distribution $\mathbb{P}_t$. The model parameters are denoted by $\theta$, and these parameters are optimized sequentially for each domain. 

Our goal is to ensure that the model can correctly segment images from previously learned domains at any given time. Specifically, we want the model parameters $\theta$ to minimize the cumulative loss from the first domain to the current domain $t_c$~\cite{buzzega2020dark}:
\begin{equation}
\arg \min _\theta \sum_{t=1}^{t_c} \mathcal{L}_t, \quad \text{ where } \mathcal{L}_t=\mathbb{E}_{(x, y) \sim D_t} \ell\left(y, f_\theta(x)\right),
\label{eq2}
\end{equation}
where $\ell$ is a non-negative sample-based loss.

In image segmentation, we consider pixel-level cross-entropy loss. That is, for each image $x \in \mathbb{P}_t$ in domain $t$, the segmentation loss $\mathcal{L}_{t}$ over all pixels $\mathcal{I}$ is given by:
\begin{equation}
\mathcal{L}_{t}\left(x, y, f_\theta\right)=-\frac{1}{|\mathcal{I}|} \sum_{i \in \mathcal{I}} \sum_{c\in|\mathcal{C}|} y_i \log \left(f_\theta^c\left(x_i\right)\right),
\label{eq3}
\end{equation}
where $y_i$ denotes the true label for pixel $x_i$. The model $f_\theta^c$ generates the probability for class $c\in \mathcal{C}$, where the set $\mathcal{C}$ includes all segmentation targets, such as foreground ($c=1$) and background ($c=0$) when $|\mathcal{C}|=2$. $|\mathcal{I}|$ represents the total number of pixels.

Optimizing Eq.~(\ref{eq2}) is particularly challenging as data from previous tasks are assumed to be unavailable. This means that the optimal configuration of $\theta$ with respect to $\mathcal{L}_{1,\cdots,t_c}$ must be found without or with little access to $D_t$ for $t\in \{1,\cdots,t_c-1\}$.
To this end, we attempt to find a parameter configuration that adapts to the current task while mimicking the output behavior for samples from previous tasks:
\begin{equation}
\small
\mathcal{L}_{t_c}+\lambda \sum_{t=1}^{t_c-1} \mathbb{E}_{x \sim \mathbb{P}_t} \left[ -\frac{1}{|\mathcal{I}|} \sum_{i \in \mathcal{I}} \sum_{c\in |\mathcal{C}|} f_{\theta_t^*}^c\left(x_i \right) \log \left(f_\theta^c\left(x_i \right)\right)  \right],
\label{eq3}
\end{equation}
where $\theta_t^*$ represents the optimal parameters learned at the end of domain $t$, and $\lambda$ is a hyperparameter balancing the trade-off of different terms.
This loss resembles the standard teacher-student knowledge distillation, encouraging the predictions of the teacher model $f_{\theta_t^*}$ to be as close as possible to those of the student model $f_{\theta}$ for each pixel $i$ in every class $c$.


To address the issue of not being able to directly access data from previous domains, we introduce a replay buffer $\mathcal{M}$ to store past experiences from all prior domains. 
Our final objective becomes:
\begin{equation}
\small
\mathcal{L}_{t_c}  +\lambda \mathbb{E}_{x \sim \mathcal{M} \cup D_{t_c} } \left[ -\frac{1}{|\mathcal{I}|} \sum_{i \in \mathcal{I}} \sum_{c\in |\mathcal{C}|} f_{\theta_t^*}^c\left(x_i \right) \log \left(f_\theta^c\left(x_i \right)\right)  \right],
\label{eq4}
\end{equation}
where $D_{t_c}$ refers to training samples from current domain $t_c$, $``\cup"$ denotes the union of two sets.

That is, we apply standard teacher-student knowledge distillation to both buffered data and samples from the current task. In practice, we use reservoir sampling~\cite{vitter1985random} to dynamically maintain a fixed-size buffer containing representative samples from prior tasks.

\subsection{Dual-Alignment Knowledge Retention} \label{sec:3.3}

We propose constructing dependencies between previous and present data and network properties to reduce catastrophic forgetting. We describe a dual-alignment strategy: Cross-Network Alignment (CNA) using linear HSIC to align old and new features (see section \ref{sec:3.3.1}) and Cross-Representation Alignment (CRA) using buffered data to align current task representations (see section \ref{sec:3.3.2}).

\subsubsection{Cross Representation Alignment}
\label{sec:3.3.1}
In our cross-representation alignment framework, the goal is to maximize the dependence between latent representations learned from both the buffered data and samples from current task. This is driven by the need to enhance the sharing of task-invariant knowledge across domains. By maximizing this dependence, we aim to facilitate better knowledge transfer and mitigate catastrophic forgetting. 

We represent the feature maps in the bottleneck layer of U-Net in a mini-batch of size $N$ as $\left\{F_{S, i}^{\mathcal{M}}\right\}_{i=1}^N$ and $\left\{F_{S, i}^{D_t}\right\}_{i=1}^N$, where $F_{S, i}^{\mathcal{M}} \in \mathbb{R}^{C \times H \times W} $ and $F_{S, i}^{D_t} \in \mathbb{R}^{C \times H \times W} $. Here, $C$ denotes the number of channels, and $H$ and $W$ are the height and width of the feature maps, respectively. A standard U-Net with 3 hidden layers is used, resulting in a latent representation per sample of dimension 128 × 12 × 12.

We intend to use HSIC with RBF kernel (Eq.~(\ref{eq:hsic_nonlinear})) to align $\left\{F_{S, i}^{\mathcal{M}}\right\}_{i=1}^N$ and $\left\{F_{S, i}^{D_t}\right\}_{i=1}^N$. However, directly applying HSIC poses two challenges. First, each representation is a 3D tensor rather than a vector. More critically, there is no explicit pairing information between $\left\{F_{S, i}^{\mathcal{M}}\right\}_{i=1}^N$ and $\left\{F_{S, i}^{D_t}\right\}_{i=1}^N$, which further complicates the alignment process. In standard HSIC (see Eq.~(\ref{eq:hsic_nonlinear})) or any existing dependence estimator, the correspondence between samples from two variables must be known. Unfortunately, this information is not available in our case. It is unclear if $F_{S,1}^{\mathcal{M}}$ should be paired with $F_{S,1}^{\mathcal{D}_t}$, $F_{S,2}^{\mathcal{D}_t}$, or another option. We thus introduce a Feature Mapping (FM) block and a Feature Pairing (FP) block to facilitate the seamless use of HSIC.
\paragraph{Feature Mapping Block} 
The FM block transforms a $3D$ feature $F$ of size $C\times H\times W$ into a vector representation of size $W\times H$ by a nonlinear function $\varphi$. In the bottleneck-layer representation of U-Net, each channel extracts certain structured information, and the impact of different channels on the final result varies. Hence, $\varphi$ takes the form of:
\begin{equation}
\varphi(F)=\frac{1}{C} \sum_{k=1}^C \lambda_k \cdot\left|F_k\right|,
\end{equation}
where $F_k$ is the feature map of the $k$-th channel, and $\lambda_k$ is the corresponding weight. 

The calculation of weight value $\lambda_k$ follows two steps. First, global average pooling is applied to the feature map of each channel to calculate the average value of the spatial features of each channel, resulting in a ``representative" value for each channel. The calculation is as follows:
\begin{equation}
Z(F_k)=\frac{1}{W \times H} \sum_{i=1}^W \sum_{j=1}^H \left|F_k(i,j)\right|,
\end{equation}
where $i$ and $j$ are the spatial positions (pixel coordinates) within the feature map. That is, $F_k(i,j)$ denotes the $(i,j)$-th pixel value in the $k$-th feature map.



Next, the softmax function is applied to normalize the distribution of channel values, yielding the weights for each channel $\lambda=\left[\lambda_1, \lambda_2, \ldots, \lambda_C\right]$:
\begin{equation}
    \lambda_k = \frac{e^{Z(F_k)}}{ \sum_{k=1}^C e^{Z(F_k)} }.
\end{equation}

\paragraph{Feature Pairing Block} 

After FM block, we obtain $\left\{\varphi(F_{S, i}^{\mathcal{M}}) \right\}_{i=1}^N$ and $\left\{ \varphi(F_{S, i}^{D_t}) \right\}_{i=1}^N$, in which $\varphi(F_{S, i}^{\mathcal{M}}) \in \mathbb{R}^d$, $\varphi(F_{S, i}^{D_t}) \in \mathbb{R}^d $, and $d=W\times H$.

Let $\pi$ be a permutation of the first $N$ natural numbers, then $\left\{\varphi(F_{S, \pi(i)}^{\mathcal{M}}) \right\}_{i=1}^N$ represents a reordering of samples $\left\{\varphi(F_{S, i}^{\mathcal{M}}) \right\}_{i=1}^N$. We evaluate HSIC values for all possible permutations:
\begin{equation}
    \text{HSIC}_\pi \left( \left\{\varphi(F_{S, \pi(i)}^{\mathcal{M}}) \right\}_{i=1}^N , \left\{ \varphi(F_{S, i}^{D_t}) \right\}_{i=1}^N \right).
\end{equation}

Among all possible permutations, we select the one that achieves the highest HSIC, i.e,
\begin{equation}
\pi^*=\arg \max _{\pi} \operatorname{HSIC}^{(\pi)}.
\label{eq29}
\end{equation}

The final dependence measure for $\left\{\varphi(F_{S, i}^{\mathcal{M}}) \right\}_{i=1}^N$ and $\left\{ \varphi(F_{S, i}^{D_t}) \right\}_{i=1}^N$, a.k.a., the regularization term used in our CRA module is given by:
\begin{equation}
    \mathcal{L}_{\text{CRA}} = - \operatorname{HSIC}^{(\pi^*)}.
\end{equation}


In this process, outliers are removed by calculating the median and median absolute deviation (MAD) of all HSIC values, with the outlier threshold set as 3 times the MAD. Values deviating from the median beyond this threshold are excluded. After filtering, the permutation with the highest HSIC value is selected, improving robustness by minimizing the impact of extreme values. 

Note that, in implementation, a small mini-batch size is typically used \cite{r2.22,zhang2023s}. Specifically, we set mini-batch size $N=4$, which results in $4!=24$ permutations, making it computationally affordable. On the other hand, aligning the latent representations of the underlying model on samples from both the buffer and current task is motivated by~\cite{wang2023dualhsic}. However, \cite{wang2023dualhsic} does not account for the ordering of samples, which leads to random pairings and significantly increases the variance or stochasticity of the results. In our experiment in Sec.~\ref{sec:4.5}, we further demonstrate that a feature pairing block is crucial for ensuring reliable performance.


\subsubsection{Cross-Network Alignment}
\label{sec:3.3.2}
 
We also align the latent representations of samples from the buffer as learned by the new network (i.e., $F_S^{\mathcal{M}}$) with those learned by the old model $F_T^{\mathcal{M}}$. Intuitively, if $F_S^{\mathcal{M}}$ is highly correlated with $F_T^{\mathcal{M}}$, the new network is expected to exhibit similar discriminative capabilities as the teacher network.

Unlike the CRA module, each sample generates representations in both the old and new networks, establishing a clear correspondence between them. Thus, the complete set of observations for alignment can be represented as $\{ (F_{S,i}^{\mathcal{M}}, F_{T,i}^{\mathcal{M}}) \}_{i=1}^N $, where \( N \) denotes the mini-batch size. This simplifies the computation of HSIC, as we only need to transform \( F_S^{\mathcal{M}} \) and \( F_T^{\mathcal{M}} \) into a vector representation. To achieve this, we concatenate all elements in \( F_S^{\mathcal{M}} \) or \( F_T^{\mathcal{M}} \) into a single vector for simplicity, resulting in \( f_S^{\mathcal{M}} \) and \( f_T^{\mathcal{M}} \), in which \( f_S^{\mathcal{M}} \in \mathbb{R}^{d'} \), \( f_T^{\mathcal{M}} \in \mathbb{R}^{d'} \), and $d'=C\times H\times W$ represents the feature dimension.

We use the linear HSIC in Eq.~(\ref{eq:hsic_linear}) herein, as it eliminates the need to tune a hyperparameter $\sigma$, the kernel size in the RBF kernel. We start by rescaling all elements in both \( f_S^{\mathcal{M}} \) and \( f_T^{\mathcal{M}} \) to the range $[0,1]$ with their respective $\ell_2$-norms:
\begin{equation}
f_T^{\mathcal{M}}= F_T^{\mathcal{M}}/ \| F_T^{\mathcal{M}} \|_2, \quad 
f_S^{\mathcal{M}}= F_S^{\mathcal{M}}/ \| F_S^{\mathcal{M}} \|_2.
\label{eq32}
\end{equation}

Next, we normalize the rescaled feature representations to have zero mean and unit variance, and construct a cross-correlation matrix $C_{st}=\frac{ (f_S^{\mathcal{M}})^{\top} f_T^{\mathcal{M}}}{N} \in \mathbb{\mathcal{R}}^{d' \times d'}$, resulting in the linear HSIC regularization as follows:
\begin{equation}
\label{eq:hsic_cma}
    \widehat{\text{HSIC}}_{\text{CNA}} = \|C_{st}\|_F^2.
\end{equation}

A trivial solution to Eq.~(\ref{eq:hsic_cma}) is $(C)_{ij}=1, \forall 1\leq i,j\leq d' $, which is not ideal since the perfect correlation $(+1)$ between different dimensions of the representations implies a low power of the representations~\cite{zbontar2021barlow}. Hence, we focus only on the diagonal entries $v_i = (C_{st})_{ii}$, as the corresponding dimensions between the two sets of representations encode similar information. 
To formulate $v_i$ close to $1$ as a minimization problem, we design the following loss function:
\begin{equation}
\mathcal{L}_{\text {CNA}}=\log _2 \sum_{i=1}^{d'} (v_i-1)^{2 \alpha}.
\label{eq34}
\end{equation}

One reason for the ``$-\log$'' is that every probability distribution can be thought of as a compression algorithm, and the negative $\log_2$ probability is the number of bits you need to encode with this compression algorithm.

\section{Experiments}
\label{sec:experiments}
\subsection{Datasets and preprocessing}
The prostate dataset~\cite{r4.1} includes T2-weighted MRI images. The training sequence (RUNMC $\rightarrow$ BMC $\rightarrow$ I2CVB $\rightarrow$ UCL $\rightarrow$ BIDMC $\rightarrow$ HK) is randomly determined. All images are adjusted to $192\times192$ in the axial plane and normalized to the range $[0, 1]$. Public datasets of fundus images from four different clinical centers \cite{r4.2} are used for optic cup and optic disc segmentation. During preprocessing, all images are adjusted to $192\times192$ in the axial plane and normalized to the range $[0, 1]$. The continual learning begins from domain 1 to domain 4. 

\begin{figure}[t]  
    \centering
    \includegraphics[width=0.75\columnwidth, trim=10 15 10 15, clip]{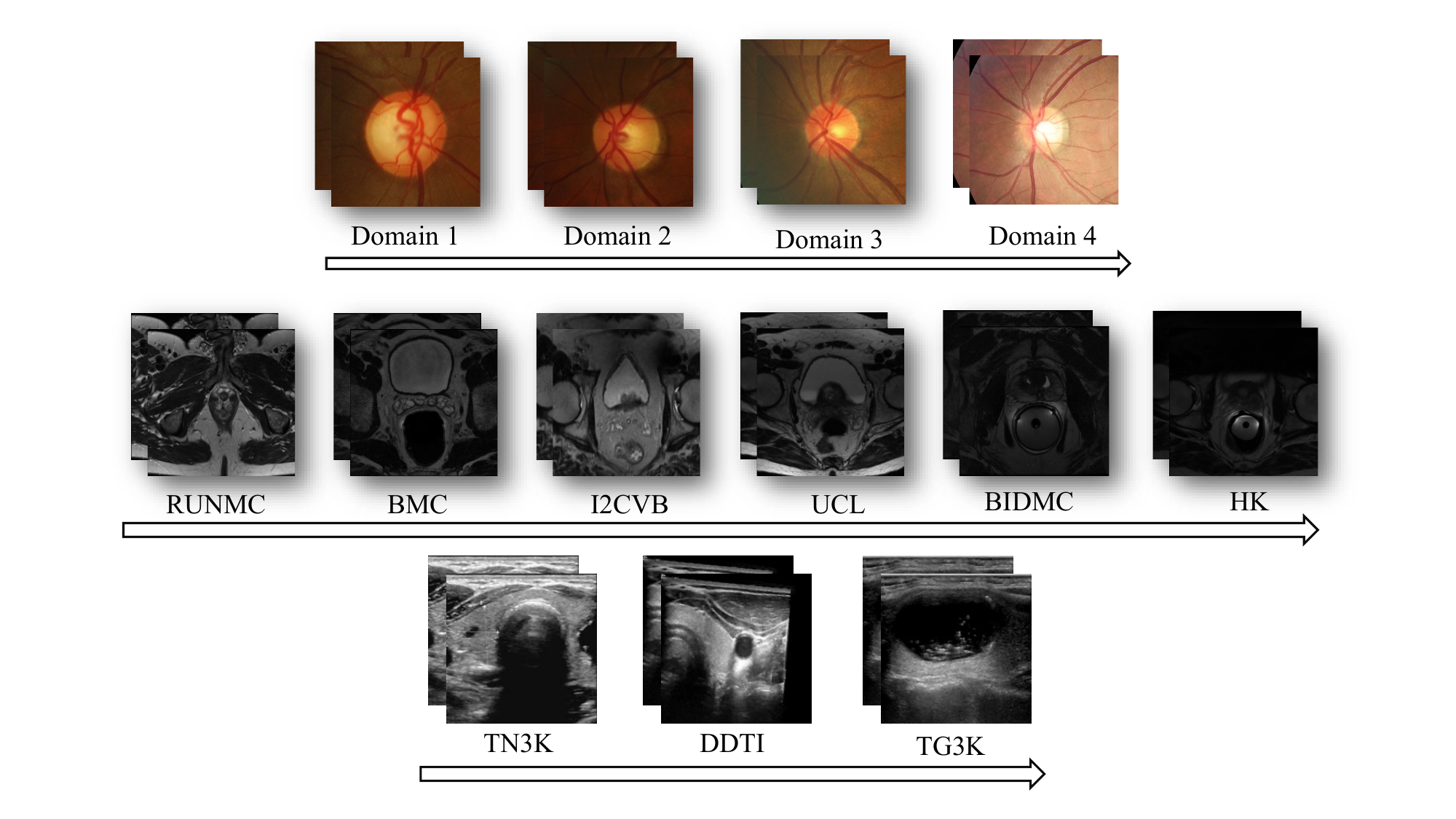}
    \caption{Data sample}
    \label{figuredatasample}
\end{figure}

Additionally, we constructed a dataset with three domains for thyroid nodules: TN3K \cite{TN3K}, TG3K \cite{TG3K}, DDTI \cite{DDTI}, ith each domain sourced from public datasets. The training order is TN3K $\rightarrow$ DDTI $\rightarrow$ TG3K. All images are resized to 192×192 in the axial plane and normalized to the range [0, 1]. Table~\ref{table_data_sample} and Fig.~\ref{figuredatasample} provide details of datasets.

\begin{table}[t]\centering
\renewcommand{\arraystretch}{1.2}
\setlength{\abovecaptionskip}{0pt}
\vspace{-0.2cm}
\caption{Datasets}
\scriptsize  
\begin{tabular}{>{\centering\arraybackslash}p{2cm}|>{\centering\arraybackslash}p{2cm}|>{\centering\arraybackslash}p{1.5cm}} 
\hline \hline
Task & Domain ID & {Number of samples} \\ \hline
\multirow{4}{*}{Fundus} & Domain 1 & 101 \\  
  & Domain 2 & 159 \\  
  & Domain 3 & 400\\  
  & Domain 4 & 400\\ \hline
\hline
& Domain ID & Case num\\ \hline
\multirow{6}{*}{Prostate} & RUNMC & 30 \\  
  & BMC   & 30 \\  
  & I2CVB & 19 \\   
  & UCL   & 13 \\  
  & BIDMC & 12 \\  
  & HK    & 12 \\ \hline
\hline
& Domain ID & Number of samples \\ \hline
\multirow{3}{*}{Thyroid nodules} & TN3K & 667 \\  
  & DDTI & 408 \\  
  & TG3K & 607 \\ \hline
\hline
\end{tabular}
\label{table_data_sample}
\end{table}

\begin{table*}[!t]
\centering
\renewcommand{\arraystretch}{1.5}
\caption{Comparative results for optic cup segmentation. We report the performance of past (Domain1, Domain2, Domain3) and current (Domain4) domains, at the end of training. AVG is the average performance on all domains. BWT shows the degree of forgetting.}
\vspace{0cm}
\resizebox{\linewidth}{!}{
\begin{tabular}{c|c|c}
\hline
\textbf{Dice Coefficient (Dice) $\% \uparrow$} & \textbf{IOU \% $\uparrow$} & \textbf{Hausdorff Distance 95 $\downarrow$} \\
\hline
\begin{tabular}{c|cccc|cc}
Task & Domain1 & Domain2 & Domain3 & Domain4 & AVG & BWT \\ \hline
Upper & 82.43 & 75.36 & 85.16 & 84.73 & 81.92 & - \\ \hline
EWC & 54.13 & \underline{62.67} & 83.28 & 86.30 & 71.59 & -14.80 \\ 
KD & \underline{58.51} & 61.36 &  \textbf{84.31} & 86.02 & \underline{72.55} & \underline{-13.28} \\ 
MAS & 53.87 & 62.78 & 83.35 & 86.43 & 71.61 & -15.20 \\ 
PLOP & 55.62 & 61.55 & 82.06 & 86.05 & 71.32 & -14.54 \\ 
MIB  & 56.58 & 60.42 & \underline{83.75} & 86.58 & 71.83 & -14.16 \\ 
SEQ & 45.45 & 55.29 & 81.95 & \underline{86.91} & 67.40 & -20.28 \\ 
TED & 57.26 & 60.99 & 83.29 &  \textbf{87.20} & 72.18 & -14.20 \\
MRSS & 54.68 & 61.31 & 83.02 & 86.74 & 71.44 & -14.52\\
Ours &  \textbf{70.44} &  \textbf{71.53} & 83.30 & 86.35 &  \textbf{77.91} &  \textbf{-5.73} \\ \hline
\end{tabular} &

\begin{tabular}{cccc|cc}
Domain1 & Domain2 & Domain3 & Domain4 & AVG & BWT \\ \hline
71.39 & 62.73 & 74.68 & 74.40 & 70.80 & - \\ \hline
39.61 & \underline{48.63} & 71.80 & 76.55 & 59.14 & -16.60 \\ 
43.31 & 47.36 &  \textbf{73.25} & 76.05 & \underline{59.99} & \underline{-15.16} \\ 
39.23 & \underline{48.36} & 71.87 & 76.67 & 59.03 & -17.08 \\ 
40.94 & 47.35 & 70.06 & 76.19 & 58.63 & -16.50 \\ 
\underline{42.08} & 46.65 & \underline{72.47} & 76.95 & 59.54 & -15.74 \\ 
39.61 & 48.63 & 71.80 & 76.55 & 59.14 & -22.16 \\ 
41.91 & 46.83 & 71.80 &  \textbf{77.87} & 59.60 & -16.32 \\ 
39.81 & 47.12 & 71.39 & \underline{77.14} & 58.87 & -16.42\\
 \textbf{55.80} &  \textbf{57.36} & 71.88 & 76.61 &  \textbf{65.41} &  \textbf{-7.48} \\ \hline
\end{tabular} &

\begin{tabular}{cccc|cc}
Domain1 & Domain2 & Domain3 & Domain4 & AVG & BWT \\ \hline
10.27 & 12.13 & 6.87 & 8.73 & 9.50 & - \\ \hline
21.11 & 14.36 & 6.70 & 4.39 & 11.64 & 2.68 \\ 
24.52 & 17.27 &  \textbf{5.80} & 4.08 & 12.92 & 4.11 \\ 
20.47 & 18.82 & 6.15 & 5.53 & 12.74 & 3.35 \\ 
24.56 & 18.57 & 7.03 &  \textbf{4.02} & 13.54 & 4.54 \\ 
22.14 & 19.25 & \underline{6.10} & \underline{4.05} & 12.88 & 3.33 \\ 
33.61 & 31.52 & 7.32 & 4.08 & 19.13 & 9.47 \\ 
\underline{19.79} & \underline{13.99} & 6.23 & 4.18 & \underline{11.05} & \underline{1.27} \\ 
20.08 & 15.06 & 6.72 & 4.52 & 11.59 & 2.71 \\
 \textbf{16.07} &  \textbf{11.58} & 6.47 & 4.06 &  \textbf{9.54} &  \textbf{-0.52} \\ \hline
\end{tabular}
\end{tabular}}
\label{table1}
\end{table*}

\begin{table*}[!t]
\renewcommand{\arraystretch}{1.5}
\centering
\caption{Comparative results for optic disk segmentation. We report the performance of past (Domain1, Domain2, Domain3) and current (Domain4) domains, at the end of training. AVG is the average performance on all domains. BWT shows the degree of forgetting.}
\vspace{0cm}
\resizebox{\linewidth}{!}{
\begin{tabular}{c|c|c}
\hline
\textbf{Dice Coefficient (Dice) $\% \uparrow$} & \textbf{IOU \% $\uparrow$} & \textbf{Hausdorff Distance 95 $\downarrow$} \\
\hline
\begin{tabular}{c|cccc|cc}
Task & Domain1 & Domain2 & Domain3 & Domain4 & AVG & BWT \\ \hline
Upper & 93.53 & 91.24 & 95.34 & 95.25 & 93.84 & - \\ \hline
EWC & 79.81 & \underline{83.41} & 93.52 & 95.39 & 88.03 & -8.89 \\ 
KD & 78.27 & 81.13 & 92.91 & 95.36 & 86.92 & -10.12 \\ 
MAS & \underline{81.14} & 83.58 & \underline{93.68} & 95.28 & 88.42 & -8.24 \\ 
PLOP & 78.75 & 81.86 & 93.27 & 95.37 & 87.31 & -9.90 \\ 
MIB & 78.29 & 82.26 & 91.69 &  \textbf{95.49} & 86.93 & -10.32 \\ 
SEQ & 77.51 & 82.46 & 93.52 & 95.36 & 87.21 & -9.88 \\ 
TED & 81.07 & 82.09 & 93.57 & \underline{95.44} & \underline{88.04} & \underline{-8.81} \\
MRSS & 77.77 & 81.91 & 93.13 & \underline{95.44} & 87.06 & -10.17 \\
Ours &  \textbf{85.76} &  \textbf{84.95} &  \textbf{93.79} & 95.21 &  \textbf{89.93} &	 \textbf{-4.68} \\ \hline
\end{tabular} &

\begin{tabular}{cccc|cc}
Domain1 & Domain2 & Domain3 & Domain4 & AVG & BWT \\ \hline
88.09 & 84.12 & 91.22 & 91.05 & 88.62 & - \\ \hline
67.16 & 72.40 & 88.00 & 91.29 & 79.71 & -13.84 \\ 
65.02 & 69.09 & 86.93 & 91.24 & 78.07 & -15.60 \\ 
\underline{68.98} & \underline{72.60} & \underline{88.27} & 91.10 & \underline{80.24} & \underline{-12.94} \\ 
65.71 & 70.11 & 87.53 & 91.26 & 78.65 & -15.35 \\ 
64.91 & 70.61 & 84.92 & \textbf{91.47} & 77.98 & -16.10 \\ 
64.13 & 70.95 & 87.99 & 91.25 & 78.58 & -15.18 \\ 
68.89 & 70.48 & 88.05 & \underline{91.38} & 79.70 & -13.74 \\
64.45 & 70.30 & 87.32 & 91.37 & 78.36 & -15.63 \\ 
\textbf{75.61} & \textbf{74.54} & \textbf{88.44} & 90.96 &  \textbf{82.39} &  \textbf{-7.45}
\\ \hline
\end{tabular} &

\begin{tabular}{cccc|cc}
Domain1 & Domain2 & Domain3 & Domain4 & AVG & BWT \\ \hline
6.88 & 6.51 & 3.83 & 2.99 & 5.05 & - \\ \hline
24.69 & 15.37 & \underline{5.60} & 2.98 & 12.16 & 9.48 \\ 
22.04 & 15.35 & 6.44 & 2.93 & 11.69 & 8.84 \\ 
22.81 & 14.16 & 5.70 & 2.86 & 11.38 & 8.21 \\ 
25.87 & 16.70 & 7.00 & 2.87 & 13.11 & 10.99 \\ 
22.60 & 14.51 & 8.55 & 2.83 & 12.12 & 9.25 \\ 
27.74 & 20.42 & 7.13 & 3.16 & 14.61 & 12.94 \\ 
\underline{19.71} & \underline{13.91} & 6.12 &  \textbf{2.82} & \underline{10.64} & \underline{7.17} \\ 
30.41 & 20.76 & 6.60 & 3.28 & 15.26 & 13.22\\
 \textbf{16.05} &  \textbf{11.88} &  \textbf{5.47} & \underline{2.92} &  \textbf{9.08} &  \textbf{3.54}
 \\ \hline
\end{tabular}
\end{tabular}}
\label{table2}
\end{table*}

\begin{table*}[!t]
\renewcommand{\arraystretch}{2.5}
\centering
\caption{Comparative results for prostate segmentation. We report the performance of past (RUNMC, BMC, I2CVB, UCL, and BIDMC) and current (HK) domains, at the end of training. AVG is the average performance on all domains. BWT shows the degree of forgetting.}
\vspace{0cm}
\renewcommand{\arraystretch}{1.5} 
\resizebox{\linewidth}{!}{
\begin{tabular}{c|c|c}
\hline
\textbf{Dice Coefficient (Dice) $\% \uparrow$} & \textbf{IOU \% $\uparrow$} & \textbf{Hausdorff Distance 95 $\downarrow$} \\
\hline
\begin{tabular}{c|cccccc|cc}
Task & RUNMC & BMC & I2CVB & UCL & BIDMC & HK & AVG & BWT \\ \hline
Upper & 88.16 & 84.73 & 74.15 & 83.96 & 67.30 & 71.60 & 78.32 & - \\ \hline
EWC & 62.28 & 60.75 & 42.71 & 68.40 & \underline{52.90} & 71.55 & 59.76 & -23.25 \\ 
KD & 73.56 & 67.10 & 51.29 & 74.14 & 39.15 & 76.39 & 63.60 & -18.72 \\ 
MAS & 68.47 & 65.76 & 45.80 & 73.07 & 42.79 & 70.07 & 60.99 & -21.37 \\ 
PLOP & 67.44 & 66.90 & 40.45 & 71.40 & 29.40 & 65.97 & 56.93 & -25.13 \\ 
MIB & 67.20 & 69.02 & 54.36 & \underline{74.90} & 34.57 & 71.75 & 61.97 & -19.16 \\ 
SEQ & \underline{75.96} & 69.04 & 48.57 & 70.53 & 63.79 & 75.80 & \underline{67.28} & \underline{-12.25} \\ 
TED & 74.31 & \underline{69.25} & \underline{55.06} & 74.24 & 47.66 & 75.86 & 66.06 & -15.71 \\ 
MRSS & 59.25 & 55.22 & 30.81 & 62.74 & 50.46 & 74.44 & 55.48 & -28.34\\ 
Ours & \textbf{86.20} & \textbf{72.63} & \textbf{67.55} &  \textbf{80.30} & \textbf{64.99} & \textbf{81.26} &  \textbf{75.49} & \textbf{-3.91} \\ \hline
\end{tabular} &

\begin{tabular}{cccccc|cc}
RUNMC & BMC & I2CVB & UCL & BIDMC & HK & AVG & BWT \\ \hline
78.98 & 73.73 & 59.78 & 72.44 & 50.80 & 55.83 & 65.26 & - \\ \hline
46.19 & 45.60 & 29.52 & 52.32 & 37.11 & 55.92 & 44.44 & -26.14 \\ 
58.74 & 52.07 & 37.20 & 59.02 & 26.43 & \underline{62.11} & 49.26 & -20.20 \\ 
52.91 & 51.36 & 32.23 & 57.68 & 28.41 & 54.55 & 46.19 & -23.52 \\ 
51.96 & 52.15 & 27.85 & 55.74 & 19.02 & 50.45 & 42.86 & -26.31 \\ 
51.61 & 54.61 & 39.92 & 60.04 & 23.51 & 56.28 & 47.66 & -20.59 \\ 
\underline{62.21} & \underline{54.77} & 33.75 & 55.07 & \underline{47.28} & 61.17 & \underline{52.37} & \underline{-14.23} \\ 
59.50 & 54.60 & \underline{40.17} & \underline{59.32} & 32.50 & 61.35 & 51.24 & -17.92 \\ 
43.10 & 39.94 & 19.33 & 46.35 & 34.15 & 59.62 & 40.41 & -30.65 
\\
 \textbf{75.89} &  \textbf{58.66} &  \textbf{52.24} &  \textbf{67.36} &  \textbf{49.24} &  \textbf{68.48} &  \textbf{61.98} &  \textbf{-4.69} \\ \hline
\end{tabular} &

\begin{tabular}{cccccc|cc}
RUNMC & BMC & I2CVB & UCL & BIDMC & HK & AVG & BWT \\ \hline
10.75 & 23.57 & 32.67 & 26.27 & 51.48 & 46.02 & 31.79 & - \\ \hline
43.25 & 40.93 & 66.36 & 45.04 & 33.79 & 39.55 & 44.82 & 30.02 \\ 
28.92 & 27.82 & 52.69 & 35.91 & 44.93 & 20.08 & 35.06 & 19.92 \\ 
26.79 & 31.27 & 55.94 & 34.30 & 56.20 & 28.84 & 38.89 & 25.51 \\ 
\underline{23.14} & 31.74 & 74.81 & 33.23 & 67.52 & 31.38 & 43.64 & 29.88 \\ 
40.33 & \underline{25.43} & \underline{44.16} & 41.22 & 60.53 & 21.55 & 38.87 & 23.16 \\ 
26.32 & 28.67 & 50.34 & 41.58 & 33.94 & 33.13 & 35.66 & 11.61 \\ 
23.75 & 28.71 & 49.62 & \underline{33.56} & 44.59 & 22.93 & \underline{33.86} & \underline{19.83} \\ 
42.53 & 42.49 & 82.37 & 46.18 & \underline{31.47} & 23.10 & 44.69 & 33.34 \\
 \textbf{11.23} &  \textbf{24.20} &  \textbf{21.34} &  \textbf{9.95} &  \textbf{19.89} &  \textbf{15.30} &  \textbf{16.99} &  \textbf{-0.17} \\ \hline
\end{tabular}
\end{tabular}}
\label{table3}
\end{table*}

\begin{table*}[!t]
\centering
 \renewcommand{\arraystretch}{1.2}
\caption{Comparative results for thyroid nodule segmentation. We report the performance of past (TN3K, DDTI, TG3K) and current (Ours) domains, at the end of training. AVG is the average performance on all domains. BWT shows the degree of forgetting.}
\vspace{0cm}
\resizebox{\linewidth}{!}{
\begin{tabular}{c|c|c}
\hline
\textbf{Dice Coefficient (Dice) $\% \uparrow$} & \textbf{IOU \% $\uparrow$} & \textbf{Hausdorff Distance 95 $\downarrow$} \\
\hline
\begin{tabular}{c|ccc|cc}
Task & TN3K & DDTI & TG3K & AVG & BWT \\
\hline
Upper & 66.56 & 67.94 & 93.97 & 76.16 & - \\
EWC & 25.89 & 43.77 & 96.90 & 55.52 & -32.53 \\
KD & 26.01 & 43.38 & 97.89 & 55.76 & -31.60 \\
MAS & 26.93 & 46.69 & 97.42 & 57.01 & -30.47 \\
PLOP & 24.71 & 39.35 & 96.96 & 53.67 & -35.02 \\
MIB & 25.36 & 39.93 & 97.09 & 54.13 & -34.70 \\
SEQ & 13.24 & 21.02 & 97.64 & 43.97 & -50.19 \\
TED & 39.17 & 54.83 & 97.97 & 63.99 & -19.69 \\
MRSS & 22.83 & 37.96 & 97.27 & 52.69 & -36.70 \\
Ours & 60.31 & 58.47 & 95.09 & 71.29 & -5.73 \\
\hline
\end{tabular}
&
\begin{tabular}{ccc|ccc}
TN3K & DDTI & TG3K & AVG & BWT \\
\hline
55.17 & 55.99 & 89.33 & 66.83 & - \\
16.48 & 30.64 & 94.52 & 47.21 & -31.59 \\
16.51 & 30.23 & 96.07 & 47.60 & -30.99 \\
17.15 & 33.09 & 95.11 & 48.45 & -29.97 \\
15.58 & 27.04 & 94.44 & 45.69 & -33.65 \\
16.16 & 27.58 & 94.68 & 46.14 & -33.32 \\
7.54 & 12.57 & 95.59 & 38.57 & -45.13 \\
27.83 & 41.51 & 96.22 & 55.19 & -20.08 \\
14.18 & 25.73 & 95.00 & 44.97 & -35.01 \\
48.93 & 46.60 & 90.94 & 62.16 & -5.42 \\
\hline
\end{tabular}
&
\begin{tabular}{ccc|ccc}
TN3K & DDTI & TG3K & AVG & BWT \\
\hline
49.54 & 34.51 & 10.14 & 31.40 & - \\
73.82 & 76.11 & 4.20 & 51.38 & 36.07 \\
72.60 & 75.71 & 2.26 & 50.19 & 37.57 \\
79.22 & 68.90 & 4.27 & 50.80 & 35.41 \\
76.55 & 75.01 & 5.03 & 52.20 & 36.76 \\
83.69 & 84.58 & 4.39 & 57.55 & 44.74 \\
85.00 & 88.36 & 5.62 & 59.66 & 50.94 \\
68.60 & 53.93 & 3.10 & 41.88 & 24.71 \\
87.03 & 93.37 & 4.37 & 61.59 & 50.86 \\
42.84 & 39.62 & 7.59 & 30.02 & 0.31 \\
\hline
\end{tabular}
\end{tabular}}
\label{table_thyroid}
\end{table*}



\begin{figure*}[t]  
    \centering
    \includegraphics[width=\textwidth, trim=45 25 20 30, clip]{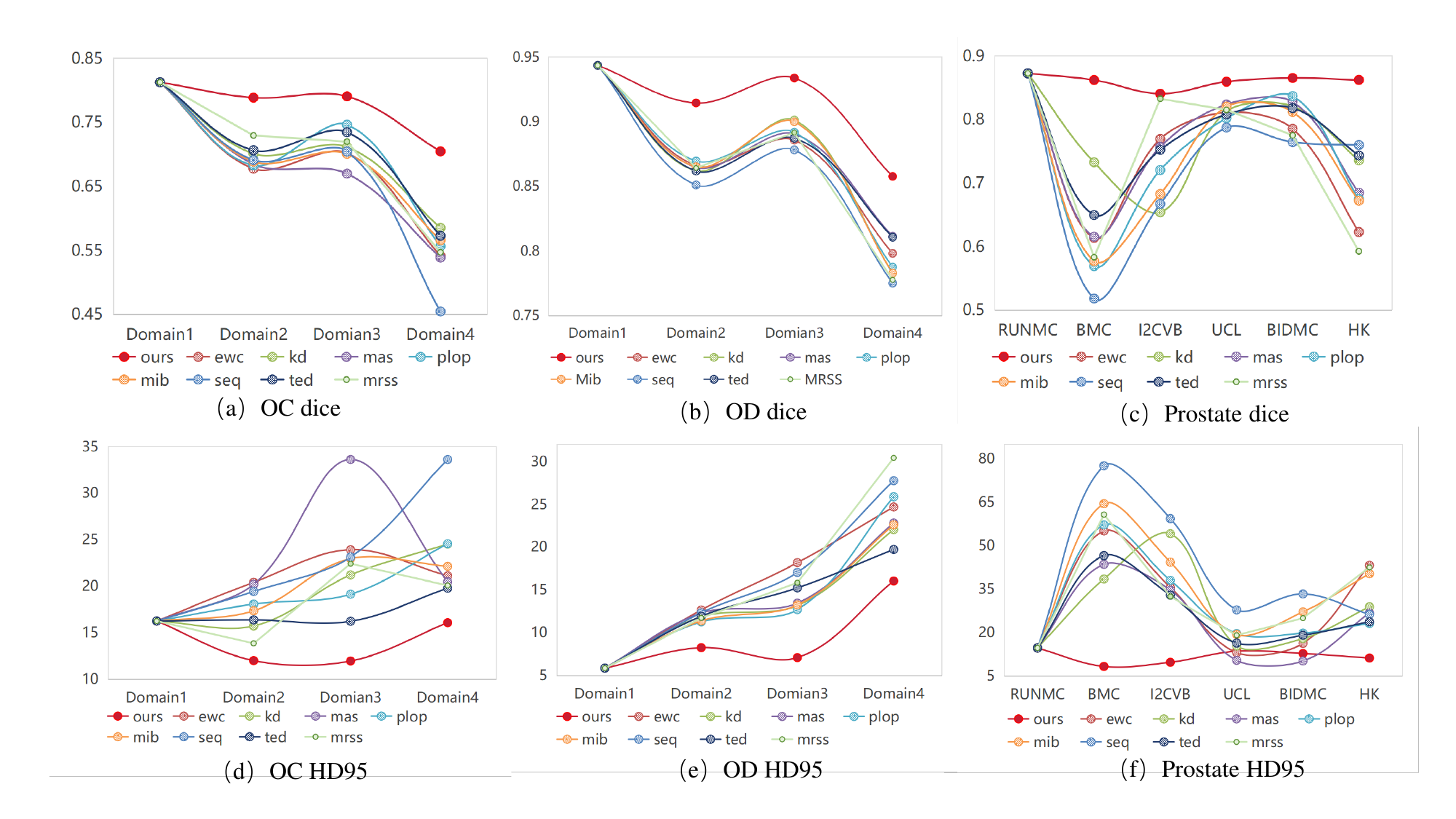}
    \caption{The forgetting curve on the first domain of the OC OD prostate dataset is presented. Since the trends of the IOU and Dice metrics are similar, we have chosen to display the Dice and HD95 metrics.}
    \label{figure2}
\end{figure*}

\begin{figure*}[t]  
    \centering
    \includegraphics[width=\textwidth, trim=45 155 45 130, clip]{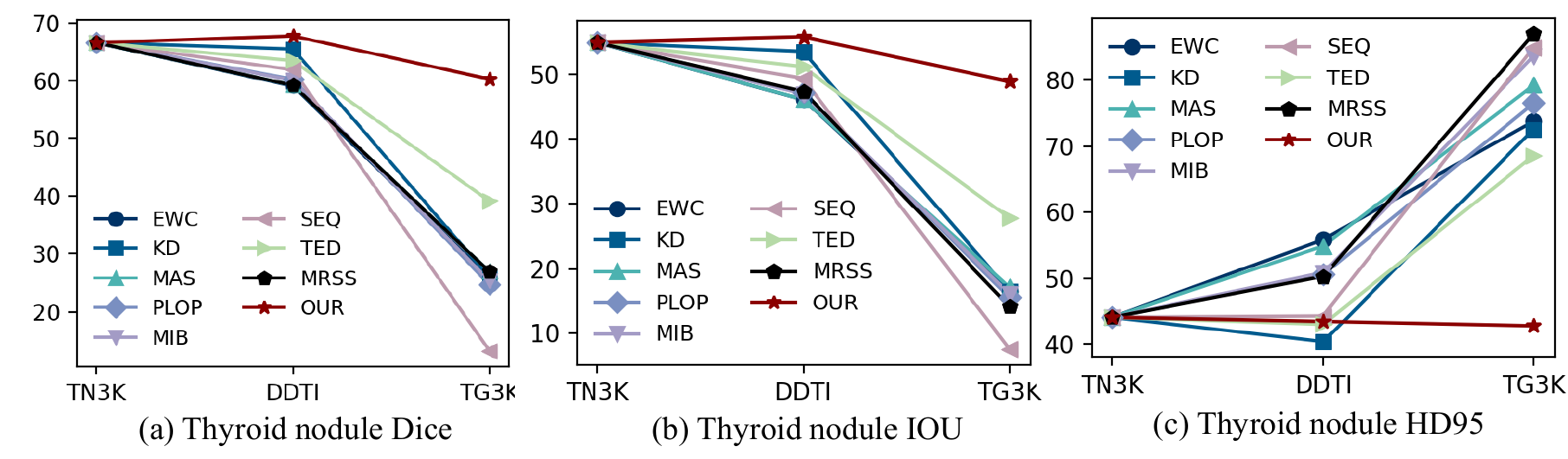}
    \vspace{-0.6cm}
    \caption{The forgetting curve on the first domain of the thyroid nodule dataset is presented.}
    \label{figure_thyroid nodule}
\end{figure*}
\begin{figure*}[t]  
    \centering
    \includegraphics[width=0.99\textwidth, trim=10 45 10 15, clip]{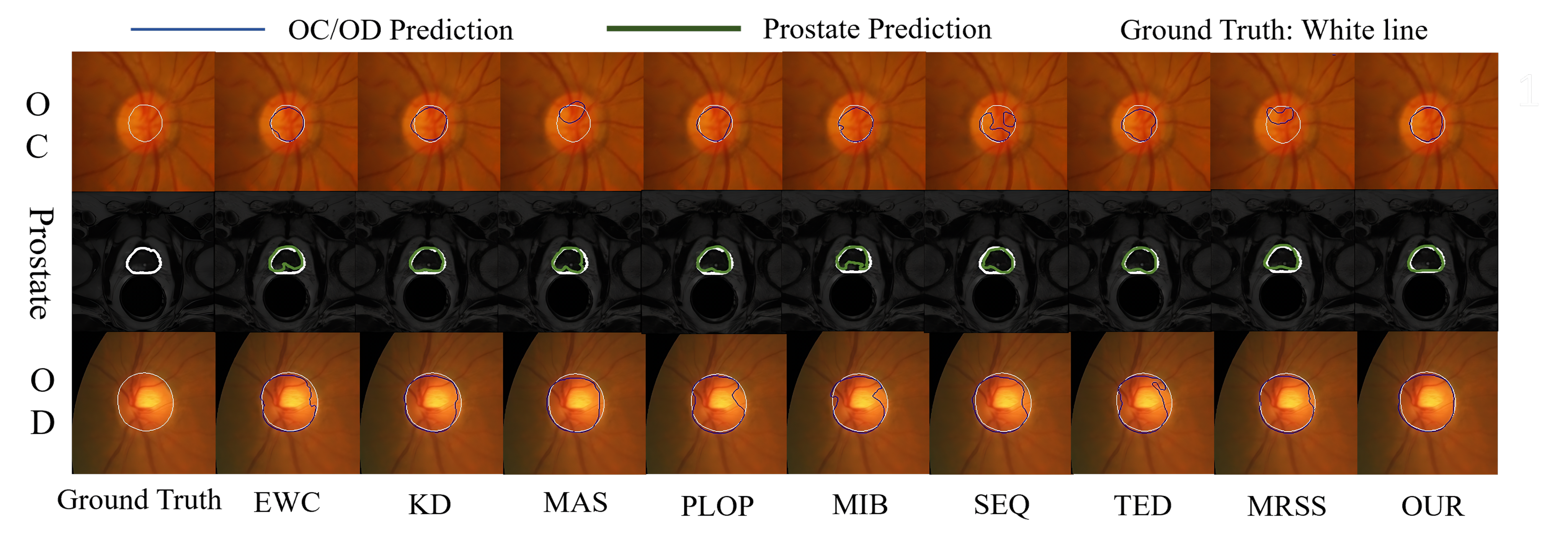}
    \caption{Segmentation visualization results, with ground truth in white. For fundus images, predictions are shown in deep blue. For prostate images, predictions in green.}
    \label{figure3}
\end{figure*}
\subsection{Experimental setting}
Our segmentation network utilizes the Mirrored Encoder-Decoder 2D-UNet architecture. Initially, we set the learning rate to 0.0002, then adjusted it to 0.0001 for subsequent datasets, and decay it at a rate of 0.99 for each training epoch. Training was conducted on a GeForce RTX 2080 GPU with 50 epochs for each dataset stage, with a batch size of 4. Data was split into training, validation, and testing sets in a ratio of 60:15:25 at each stage, and the best performing model on the validation set was selected for testing. 

Replay-Enhanced Knowledge Distillation (REKD) used a buffer of 50. For comparison methods, we used the same configuration as our method. For the Optic Cup dataset, Loss CRA has a weight of -0.75 (sigma = 0.001), and Loss CNA has a weight of 0.9 (alpha = 2). For the Optic Disk dataset, CRA remains the same, but CNA uses alpha = 1.3. For the Prostate dataset, CRA parameters are unchanged, with CNA using alpha = 1.5. Loss REKD consistently has a weight of 0.01 across all datasets. The parameters of the thyroid nodule dataset are the same as those of the  Optic Cup dataset.

\subsection{Evaluation Metrics}
We used Dice Similarity Coefficient (DSC), Intersection over Union (IoU), 95\% Hausdorff Distance (HD95), Average Accuracy (AVG), and Backward Transfer (BWT) for evaluation. Higher DSC and IoU values indicate better performance, while lower HD95 values are preferable. AVG reflects overall accuracy, and BWT measures knowledge retention, with higher BWT values being better for DSC and IoU, and lower BWT values being better for HD95.

The Dice Similarity Coefficient (DSC) measures the similarity between two sets, which is commonly used in image segmentation. It is defined as:
\begin{equation}
D S C=\frac{2|A \cap B|}{|A|+|B|},
\label{eq41}
\end{equation}
where $A$ represents the set of pixels in the ground truth and $B$ the set of pixels in the predicted segmentation.

The Intersection over Union (IoU) quantifies the overlap between two sets. It is defined as:
\begin{equation}
I o U=\frac{|A \cap B|}{|A \cup B|}.
\label{eq42}
\end{equation}

The 95\% Hausdorff Distance (HD95) measures the distance between two point sets, often used to evaluate segmentation boundaries. It is defined as:
\begin{equation}
\begin{aligned}
HD_{95}(A, B) &= \max \left\{ h_{95}(A, B), h_{95}(B, A) \right\}, \\
h_{95}(A, B) &= \max_{a \in A} \min_{b \in B} d(a, b). \\
h_{95}(B, A) &= \max_{b \in B} \min_{a \in A} d(a, b),
\end{aligned}
\end{equation}
where $\max \min $ represents the maximum of the minimum distances from each point. The 95th percentile is the distance below which 95

For DSC, backward transfer (BWT) quantifies the degree of forgetting, defined as:
\begin{equation}
B W T_{D S C}=\frac{1}{K-1} \sum_{i=1}^{K-1}\left(D S C_{K, i}-D S C_{i, i}\right).
\label{eq44}
\end{equation}

For IoU, backward transfer (BWT) is defined as:
\begin{equation}
B W T_{\mathrm{IoU}}=\frac{1}{K-1} \sum_{i=1}^{K-1}\left(I o U_{K, i}-I o U_{i, i}\right).
\label{eq45}
\end{equation}

For HD95, backward transfer (BWT) is defined as:
\begin{equation}
B W T_{H D 95}=\frac{1}{K-1} \sum_{i=1}^{K-1}\left(H D 95_{K, i}-H D 95_{i, i}\right).
\label{eq46}
\end{equation}

In terms of performance interpretation, higher values of $A V G$ and $B W T$ indicate better performance for DSC and IoU metrics, while lower values of $A V G$ and $B W T$ indicate better performance for the HD95 metric.

AVG represents the mean DSC, IOU and HD95 across all test datasets and is given by:
\begin{equation}
\mathrm{AVG}_{\text {metric }}=\frac{1}{K} \sum_{i=1}^K \operatorname{metric}_{K, i},
\end{equation}
where metric represents measures DSC, IoU, or HD95.


\begin{table}[!t]    
    \centering    
    \caption{Ablation study on the impact of REKD, CRA, and CNA in optic cup segmentation (evaluated using Dice and HD95 metrics), $\circ$ implies NO feature pairing inside CRA.}    
    \resizebox{0.6\linewidth}{!}{ 
        \tabcolsep=.05cm 
        \scriptsize 
        \renewcommand{\arraystretch}{1.5} 
        \begin{tabular}{c c c c|c|c|c|c|c|c}             
            \hline            
            & & & & \multicolumn{6}{|c}{Optic Cup Segmentation} \\             
            \hline            
            & & & & \multicolumn{6}{|c}{Dice / HD95} \\             
            \hline            
            REKD & CRA & CNA &  & Domain1 & Domain2 & Domain3 & Domain4 & AVG & BWT \\             
            \hline            
            $\checkmark$ &  &  &  & 67.14 / 19.99 & 65.87 / 24.31 & 84.00 / 6.51 & 86.93 / 3.82 & 75.99 / 13.66 & -8.85 / 6.02  \\            
            $\checkmark$ & $\circ$ & &  & 67.02 / 15.81  & 70.84 /11.49 & 82.75 / 6.87 & 86.38 / 3.91  & 76.75 / 9.52 & -7.44 / 0.40\\            
            $\checkmark$ & $\checkmark$ &  &  & 66.70 / 15.80 & 71.44 / 12.19 & 83.23 / 6.67 & 86.73 / 3.78 & 77.02 / 9.61 & -7.09 / -0.13 \\            
            $\checkmark$ & & $\checkmark$ &  & 69.26 / 16.82  & 71.03 / 10.60  & 82.70 / 6.88 & 86.46 / 4.10 & 77.36 / 9.60  &  -6.54 / -0.14 \\            
            $\checkmark$ & $\circ$ & $\checkmark$ &  & 69.04 / 16.96 & 71.47 / 11.73 & 83.47 / 6.52 & 86.23 / 4.34  & 77.55 / 9.89 & -6.06 / -0.14\\            
            $\checkmark$ & $\checkmark$ & $\checkmark$ &  & 70.44 / 16.07 & 71.53 / 11.58 & 83.30 / 6.47 & 86.35 / 4.06 & 77.91 / 9.54 & -5.73 / -0.52 \\            
            \hline        
        \end{tabular}    
    }    
    \label{table4}
\end{table}

\begin{table*}[t]
\centering
\caption{Ablation study on the impact of REKD, CRA, and CNA in optic disc segmentation (evaluated using Dice and IOU metrics). $\circ$ implies NO feature pairing inside CRA.}
\renewcommand{\arraystretch}{2}
\resizebox{\textwidth}{!}{
\begin{tabular}{c c c|c c c c c c|c c c c c c|c c c c c c}
\hline
\hline
 & & & \multicolumn{18}{|c}{Optic Disc Segmentation} \\
\hline
 & & & \multicolumn{6}{|c }{Dice Coefficient (Dice) \% $\uparrow$} & \multicolumn{6}{|c }{IOU \% $\uparrow$} & \multicolumn{6}{|c}{Hausdorff Distance 95 (HD95) \% $\downarrow$} \\
\hline
REKD & CRA & CMA & Domain1 & Domain2 & Domain3 & Domain4 & AVG & BWT & Domain1 & Domain2 & Domain3 & Domain4 & AVG & BWT & Domain1 & Domain2 & Domain3 & Domain4 & AVG & BWT \\
\hline
$\checkmark$ & & & 82.85 & 83.73 & 92.62 & 95.33 & 88.63 & -7.72 & 71.27 & 72.70 & 86.48 & 91.18 & 80.41 & -12.25 & 18.89 & 13.71 & 7.08 & 2.96 & 10.66 & 7.68 \\
$\checkmark$ & $\circ$ & & 85.01 & 83.80 & 93.07 & 95.24 & 89.28 & -6.82 & 74.44 & 72.90 & 87.21 & 91.02 & 81.39 & -10.87 & 17.22 & 13.46 & 6.36 & 2.87 & 9.98 & 6.73 \\
$\checkmark$ & $\checkmark$ & & 84.75 & 84.63 & 93.48 & 95.05 & 89.48 & -6.47 & 74.16 & 74.01 & 87.90 & 90.69 & 81.69 & -10.33 & 16.74 & 12.02 & 5.66 & 3.11 & 9.38 & 5.94 \\
$\checkmark$ & $\checkmark$ & $\checkmark$ & 85.76 & 84.95 & 93.79 & 95.21 & 89.93 & -4.68 & 75.61 & 75.54 & 88.44 & 90.96 & 82.39 & -7.45 & 16.05 & 11.88 & 5.47 & 2.92 & 9.08 & 3.54 \\
\hline
\hline
\end{tabular}}
\label{table_ODab}
\end{table*}

\begin{table*}[t]
\centering
\caption{Ablation study on the impact of REKD, CRA, and CNA in prostate segmentation (evaluated using Dice, IOU, and HD95 metrics). $\circ$ implies NO feature pairing inside CRA.}
\renewcommand{\arraystretch}{2}
\resizebox{\linewidth}{!}{
\tabcolsep=.1cm
\scriptsize
\renewcommand{\arraystretch}{1.5}
\begin{tabular}{c c c | c c c c c c c c | c c c c c c c c | c c c c c c c c}
\hline
\hline
 & & & \multicolumn{24}{c}{Prostate Segmentation} \\
\hline
 & & & \multicolumn{8}{c|}{Dice Coefficient (Dice) \% $\uparrow$} & \multicolumn{8}{c|}{IOU \% $\uparrow$} & \multicolumn{8}{c}{Hausdorff Distance 95 (HD95) \% $\downarrow$} \\
\hline
REKD & CRA & CMA & RUNMC & BMC & I2CVB & UCL & BIDMC & HK & AVG & BWT & RUNMC & BMC & I2CVB & UCL & BIDMC & HK & AVG & BWT & RUNMC & BMC & I2CVB & UCL & BIDMC & HK & AVG & BWT \\
\hline
$\checkmark$ &  &  & 81.26 & 76.18 & 58.34 & 77.88 & 63.81 & 80.50 & 72.99 & -9.62 & 69.01 & 62.60 & 44.45 & 63.86 & 48.22 & 67.44 & 59.27 & -11.39 & 21.22 & 17.13 & 36.13 & 22.09 & 26.97 & 13.00 & 22.76 & 7.15 \\
$\checkmark$ & $\circ$ &  & 85.14 & 75.22 & 55.35 & 81.57 & 65.85 & 81.65 & 74.13 & -5.56 & 74.37 & 61.52 & 40.03 & 69.05 & 49.66 & 68.99 & 60.60 & -6.41 & 14.01 & 20.72 & 57.99 & 16.67 & 18.55 & 18.08 & 24.34 & 7.71 \\
$\checkmark$ & $\circ$ & $\checkmark$ & 86.26 & 74.27 & 62.44 & 79.77 & 66.20 & 80.29 & 74.87 & -2.62 & 75.99 & 60.61 & 47.63 & 66.57 & 50.21 & 67.09 & 61.35 & -3.42 & 12.65 & 26.31 & 33.17 & 16.96 & 16.76 & 16.35 & 20.37 & 0.64 \\
$\checkmark$ & $\checkmark$ & $\checkmark$ & 86.20 & 72.63 & 67.55 & 80.30 & 64.99 & 81.26 & 75.49 & -3.91 & 75.89 & 58.66 & 52.24 & 67.36 & 49.24 & 68.48 & 61.98 & -4.69 & 11.23 & 24.20 & 21.34 & 9.95 & 19.89 & 15.30 & 16.99 & -0.17 \\
\hline
\hline
\end{tabular}
}
\label{table_prostateab}
\end{table*}

\begin{table*}[t]
\centering
\caption{Layer Ablation Experiment on the OC Dataset.}
\renewcommand{\arraystretch}{2} 
\setlength{\abovecaptionskip}{10pt} 
\resizebox{\textwidth}{!}{
\begin{tabular}{c| c c c c c c| c c c c c c| c c c c c c}
\hline\hline
& \multicolumn{6}{|c}{Dice Coefficient (Dice) \% $\uparrow$} & \multicolumn{6}{|c}{IOU \% $\uparrow$} & \multicolumn{6}{|c}{Hausdorff Distance 95 (HD95) \% $\downarrow$} \\
\hline
Layer & Domain1 & Domain2 & Domain3 & Domain4 & AVG & BWT & Domain1 & Domain2 & Domain3 & Domain4 & AVG & BWT & Domain1 & Domain2 & Domain3 & Domain4 & AVG & BWT \\
\hline
A & 70.44 & 71.53 & 83.30 & 86.35 & 77.91 & -5.73 & 55.80 & 57.36 & 71.88 & 76.61 & 65.41 & -7.48 & 16.07 & 11.58 & 6.47 & 4.06 & 9.54 & -0.52 \\
B & 68.58 & 70.01 & 84.42 & 87.08 & 77.52 & -5.18 & 53.61 & 55.97 & 73.53 & 77.61 & 65.18 & -6.61 & 15.78 & 11.38 & 5.80 & 3.69 & 9.16 & -1.21 \\
C & 70.81 & 70.64 & 84.83 & 86.88 & 78.29 & -4.19 & 56.17 & 56.68 & 74.26 & 77.34 & 66.11 & -5.44 & 14.16 & 10.64 & 6.08 & 3.78 & 8.66 & -1.87 \\
\hline\hline
\end{tabular}}
\label{table_OC_layer}
\end{table*}

\begin{table*}[t]
\centering
\caption{Layer Ablation Experiment on the OD Dataset.}
\renewcommand{\arraystretch}{2} 
\setlength{\abovecaptionskip}{10pt} 
\resizebox{\textwidth}{!}{
\begin{tabular}{c| c c c c c c| c c c c c c| c c c c c c}
\hline
\hline
& \multicolumn{6}{|c|}{Dice Coefficient (Dice) \% $\uparrow$} & \multicolumn{6}{|c|}{IOU \% $\uparrow$} & \multicolumn{6}{|c|}{Hausdorff Distance 95 (HD95) \% $\downarrow$} \\
\hline
Layer & Domain1 & Domain2 & Domain3 & Domain4 & AVG & BWT & Domain1 & Domain2 & Domain3 & Domain4 & AVG & BWT & Domain1 & Domain2 & Domain3 & Domain4 & AVG & BWT \\
\hline
A & 85.76 & 84.95 & 93.79 & 95.21 & 89.93 & -4.68 & 75.61 & 74.54 & 88.44 & 90.96 & 82.39 & -7.45 & 16.05 & 11.88 & 5.47 & 2.92 & 9.08 & 3.54 \\
B & 86.82 & 86.84 & 93.37 & 95.30 & 90.58 & -4.86 & 77.19 & 77.36 & 87.72 & 91.14 & 83.35 & -7.88 & 15.34 & 10.35 & 5.51 & 2.83 & 8.51 & 4.77 \\
C & 87.46 & 89.14 & 93.19 & 95.28 & 91.27 & -4.07 & 78.13 & 80.96 & 87.40 & 91.09 & 84.40 & -6.72 & 11.57 & 8.75 & 5.18 & 2.86 & 7.09 & 3.08 \\
\hline
\hline
\end{tabular}}
\label{table_OD_layer}
\end{table*}

\begin{table*}[t]
\centering
\caption{Layer Ablation Experiment on the Prostate Dataset.}
\renewcommand{\arraystretch}{2} 
\setlength{\abovecaptionskip}{10pt} 
\resizebox{\textwidth}{!}{
\begin{tabular}{c| c c c c c c c c| c c c c c c c c| c c c c c c c c}
\hline
\hline
& \multicolumn{8}{|c|}{Dice Coefficient (Dice) \% $\uparrow$} & \multicolumn{8}{|c|}{IOU \% $\uparrow$} & \multicolumn{8}{|c|}{Hausdorff Distance 95 (HD95) \% $\downarrow$} \\
\hline
Layer & RUNMC & BMC & I2CVB & UCL & BIDMC & HK & AVG & BWT & RUNMC & BMC & I2CVB & UCL & BIDMC & HK & AVG & BWT & RUNMC & BMC & I2CVB & UCL & BIDMC & HK & AVG & BWT \\
\hline
A & 86.20 & 72.63 & 67.55 & 80.30 & 64.99 & 81.26 & 75.49 & -3.91 & 75.89 & 58.66 & 52.24 & 67.36 & 49.24 & 68.48 & 61.98 & -4.69 & 11.23 & 24.20 & 21.34 & 9.95 & 19.89 & 15.30 & 16.99 & -0.17 \\
B & 85.01 & 75.46 & 57.85 & 81.29 & 65.24 & 80.26 & 74.18 & -4.99 & 74.23 & 61.61 & 42.96 & 68.53 & 49.55 & 67.04 & 60.65 & -5.54 & 12.30 & 20.25 & 38.69 & 20.85 & 19.92 & 13.58 & 20.93 & 2.36 \\
C & 84.73 & 77.04 & 54.77 & 79.32 & 67.90 & 78.33 & 73.68 & -4.69 & 73.68 & 63.52 & 39.96 & 65.85 & 51.73 & 64.39 & 59.85 & -5.49 & 17.70 & 19.10 & 47.79 & 21.36 & 19.82 & 22.29 & 24.68 & 6.48 \\
\hline
\hline
\end{tabular}}
\label{table_prostate_layer}
\end{table*}

\begin{figure}[t]
    \centering
    \begin{minipage}{0.45\textwidth}
        \centering
        \includegraphics[width=0.99\textwidth, trim=10 20 10 0, clip]{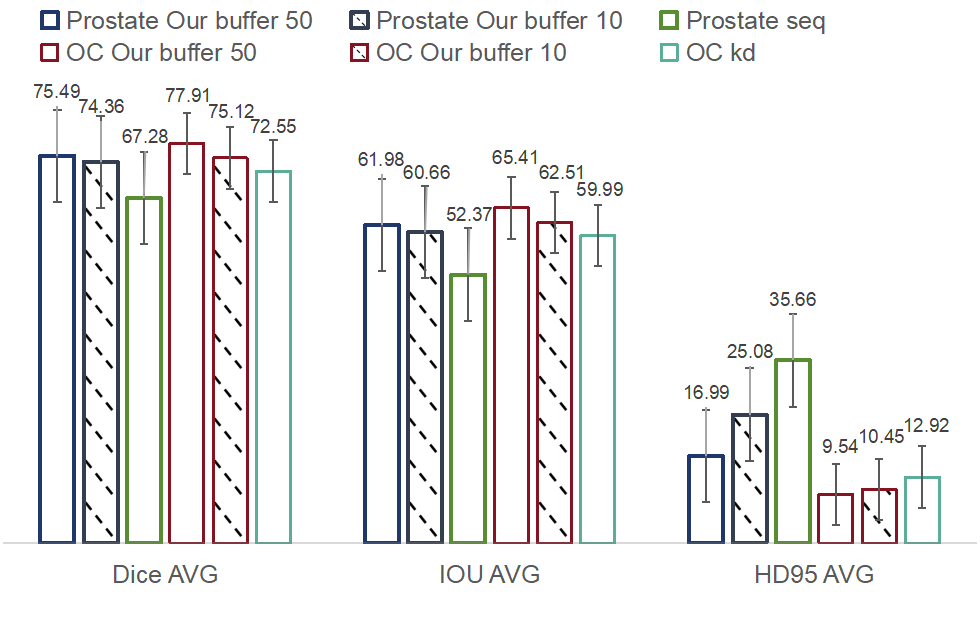}
        \caption{Buffer Ablation experiment, when our method uses a smaller buffer size of 10, it performs well, surpassing our best comparative method(seq and kd) on the Prostate and OC dataset.}
        \label{figure4}
    \end{minipage}\hfill
    \begin{minipage}{0.45\textwidth}
        \centering
        \includegraphics[width=0.99\textwidth, trim=28 2 28 50, clip]{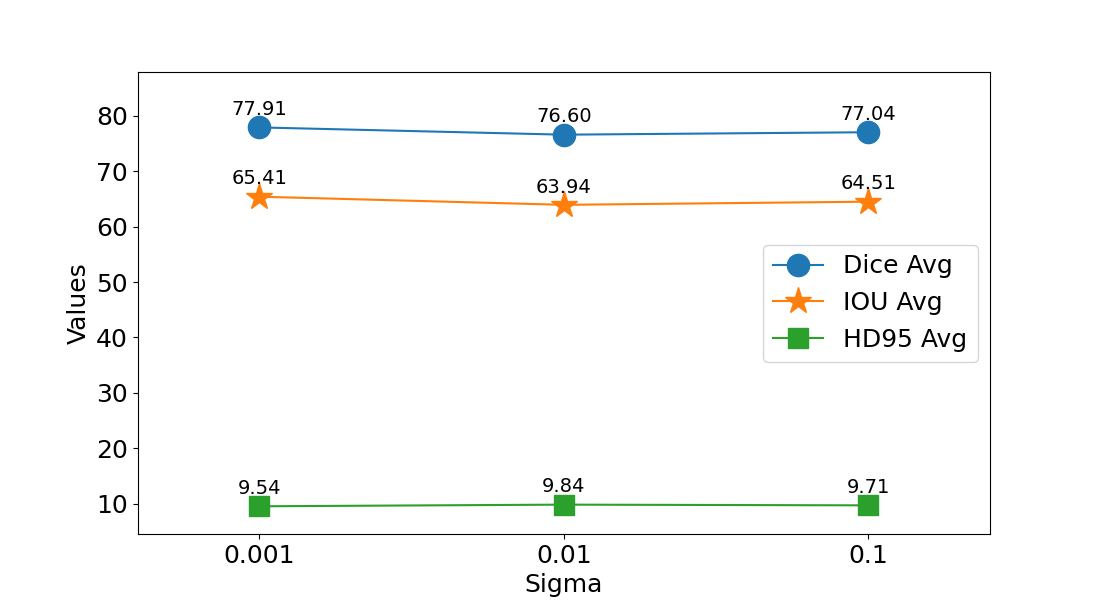}
        \caption{In CRA, the parameter sigma controls the Gaussian kernel's smoothness, which in HSIC measures independence between random variables. Experiments on OC with sigma values of 0.1, 0.01, and 0.001 show that the model remains stable with consistent performance despite changes to the kernel parameter.}
        \label{figure5}
    \end{minipage}
\end{figure}

\begin{table}[t]
\centering
\caption{Comparison of Alignment Strategies in Continual Learning (CL)} 
\label{tab:alignment_strategies} 
\resizebox{0.6\columnwidth}{!}{
\renewcommand{\arraystretch}{1.5} 
\begin{tabular}{c|c|c}
\hline
CL Strategy Type & Network Level Alignment & Feature Level Alignment \\
\hline
Replay-based & $\times$ & $\times$ \\
\hline
Regularization & $\checkmark$ & $\times$ \\
\hline
Parameter Isolation & $\times$ & $\times$ \\
\hline
Knowledge Distillation & $\times$ & $\checkmark$ \\
\hline
Ours & $\checkmark$ & $\checkmark$ \\
\hline
\end{tabular}
}
\end{table}

\subsection{Experimental result}
Our comparison methods include the upper bound performance provided by a model jointly trained on all domain datasets (Upper bound), the sequential fine-tuning (SEQ) method, and two classic regularization-based methods, EWC \cite{r4.3} and MAS \cite{r4.4}. Additionally, knowledge distillation (KD) \cite{r4.5}, MiB \cite{r4.6}, PLOP \cite{r4.7} and TED \cite{r2.22}. And there is a MRSS \cite{r2.14} based on Memory Replay. All comparison methods used the same settings as our method.

We first demonstrate the segmentation results on all domain at the end of model training in Tables~\ref{table1}-\ref{table_thyroid}. For disc segmentation, our model outperforms most baseline methods across key metrics such as Dice coefficient, IOU, and Hausdorff distance, particularly excelling in accuracy and boundary control, while exhibiting a low degree of forgetting ($\text{BWT}=-4.68$). Similarly, for prostate and thyroid nodule segmentation, our model performs exceptionally well. These findings highlight the advantages of our model in maintaining accuracy and stability across multiple domains. 
 
We then plot the forgetting curve for the first domain in Fig.~\ref{figure2} to illustrate the model's ability to retain knowledge over time. Using Dice and HD95 metrics, we observe minimal forgetting, as indicated by stable performance on the first domain despite continually learning new tasks. This result suggests our model effectively mitigates catastrophic forgetting, maintaining accuracy in earlier domains. Fig.~\ref{figure_thyroid nodule} shows the forgetting curve on the first domain of the thyroid nodule dataset. our method performs significantly better than others, with the main difference being in the forgetting on the first domain. Our method shows minimal forgetting on the first dataset and remains more stable in addressing forgetting, likely due to its ability to capture complex task dependencies, ensuring stable knowledge retention.

The segmentation results in Fig.~\ref{figure3} for both fundus and prostate images highlight our model’s strong performance in accurately identifying key anatomical structures. For fundus, it precisely segments the optic disc and cup, while in prostate, it precisely captures the prostate boundaries.

\subsection{Ablation study}\label{sec:4.5}

In Table~\ref{table4}, we demonstrate the impact of individual modules in our DAKR framework. As shown, both the CRA and CNA modules improve the performance of the baseline KD, with the best performance achieved when all regularizations are available. Furthermore, the CRA module without the feature pairing block shows a noticeable performance drop. Table \ref{table_ODab} compares OD segmentation results using Dice, IOU, and HD95 metrics across different feature combinations. As more features are incorporated, segmentation performance improves, with IOU and HD95 showing higher accuracy and lower error, especially across domains. Table \ref{table_prostateab} examines prostate segmentation, where combining CRA and CNA improves Dice and IOU scores, demonstrating the model's robustness in complex structures like the prostate.

We then demonstrate in Fig.~\ref{figure4} that our method maintains superior performance even with a small buffer size, e.g., $10$. For the Prostate dataset, it achieves a Dice score of 74.36, IOU of 60.66, and HD95 of 25.08, surpassing the best comparative method seq. For the OC dataset, it achieves a Dice score of 75.12, IOU of 62.51, and HD95 of 10.45, outperforming the best comparative method kd. 

After that, we demonstrate that our performance is insensitive to variations in the kernel size $\sigma$ of the RBF kernel when calculating the nonlinear HSIC value in the CRA module. The results in Fig.~\ref{figure5} show minimal variation in Dice scores (77.91 to 76.60), IOU (65.41 to 63.94), and HD95 (9.54 to 9.84), when $\sigma$ varies among $0.1$, $0.01$, and $0.001$.

Then, this experiment compared three different feature extraction layers: A (bottleneck layer), B (single intermediate layer, i.e., the middle layer of the encoder in U-Net), and C (combination of intermediate layers and bottleneck layer), and evaluated their performance across multiple datasets. As shown in Table~\ref{table_OC_layer} and ~\ref{table_OD_layer}, indicate that approach C outperforms A and B in most metrics, particularly in OC and OD tasks, where C demonstrates higher stability and better performance in Dice, IoU, and HD95, with significant improvement seen in retinal images (OC, OD). This is likely due to the multi-layer combination's ability to capture structural and semantic information at different scales, enhancing segmentation accuracy.

However, in the Prostate task, C did not maintain superior performance, and some metrics (e.g., HD95) even declined, as shown in Table~\ref{table_prostate_layer}. This could be related to the complexity and boundary blurriness of prostate images, where information fusion may introduce redundancy or conflicts, impacting the model’s generalization ability. These results suggest that our layer-wise alignment strategy can be extended beyond the bottleneck layer to multiple layers, although the bottleneck layer remains a reliable and consistent choice across all datasets. However, the choice of layers plays a crucial role. Future work should therefore focus on optimizing multi-layer combination strategies, potentially by incorporating attention mechanisms or adaptive layer selection methods tailored to the characteristics of different datasets.

\section{Discussion}
\label{sec:Discussion}
This study presents an innovative continual learning framework that introduces a dual alignment strategy, including Cross-Network Alignment (CNA) and Cross-Representation Alignment (CRA), effectively mitigating the catastrophic forgetting issue in continual learning. Compared with existing methods, our framework achieves strong results across various datasets, particularly in terms of model stability and knowledge transfer. As shown in Table~\ref{tab:alignment_strategies}, the replay-based method alleviates forgetting by mixing historical task data during training but lacks explicit alignment at both the network and feature levels. The regularization method prevents the destruction of historical knowledge by limiting updates to important parameters. Although it achieves some alignment at the network level, it lacks consistency modeling of the feature space. The parameter isolation method assigns separate network parameters to each task to avoid interference, but it does not consider knowledge sharing between tasks and lacks feature-level alignment. The knowledge distillation method guides the student model through the teacher model, preserving feature transfer while lacking alignment at the network level.

The innovation of this study lies in its structured approach to continual learning, considering multiple aspects of task relationships and improving the connection between current and past tasks. This approach offers a new perspective on addressing knowledge sharing and complex dependencies between tasks in continual learning.

The current framework still has certain limitations, with existing research primarily focusing on 2D data. Future work may expand to 3D scenes, addressing more complex spatial structure problems. 3D data exhibits higher-dimensional spatial topological characteristics, and there is potential for introducing point cloud or voxel-based structural alignment strategies. This could be explored in conjunction with Spatial Transformer Networks (STN) to achieve shape invariance, while leveraging Graph Attention Networks (GAT) to model the heterogeneous graph structure of 3D features, potentially capturing local and global geometric relationships and enhancing alignment robustness and memory retention across tasks. Additionally, there may be potential in constructing cross-domain models for 2D and 3D, such as 2D to 3D and 3D to 2D, to break dimensional barriers. A unified 2D-3D semantic representation in a shared task layer could broaden the model's applicability, and learning the mapping relationships between different dimensions may facilitate efficient transfer and knowledge sharing.

\section{Conclusion}
\label{sec:Conclusion}
We design a dual-alignment framework coupled with Cross-Network Alignment and Cross-Representation Alignment modules to establish dependencies between current and past tasks, thereby alleviating forgetting. Our method outperforms $8$ state-of-the-art approaches in segmentation accuracy and minimizing forgetting. It maintains stable performance even with a small buffer size. The effectiveness of both alignments, including the feature pairing block, is justified and insensitive to hyperparameters. Future work will focus on optimizing multi-layer combination strategies, exploring alignment methods for 3D data, and developing cross-domain models for 2D and 3D to enhance model applicability and knowledge transfer.

\bibliographystyle{IEEEtran}
\bibliography{main}  
\end{document}